# Spatio-temporal migration of antiferromagnetic domain walls in $Sr_2IrO_4$

Ian Robinson[1,2], David Yang[1], Ross Harder[3], Dina Sheyfer[3], Longlong Wu[1,4], Jack Griffiths[1], Emil Bozin[1,5], Mark P. M. Dean[1], Jialun Liu[2], Hengdi Zhao[6], Gang Cao[6], Angel Rodriguez-Fernandez[7], Jan-Etienne Pudell[7], Roman Shayduk[7] James Wrigley[7], Alexey Zozulya[7], Rustam Rysov[7], Aliaksandr Leonau[7,8], Ulrike Boesenberg[7], Joerg Hallmann[7] and Anders Madsen[7]

1. Condensed Matter Physics and Materials Science Division, Brookhaven National Laboratory, Upton NY 11973, USA
2. London Centre for Nanotechnology, University College, London, WC1E 6BT, UK
3. Advanced Photon Source, Argonne National Laboratory, Lemont IL 60439, USA
4. Shanghai Advanced Research Institute, Chinese Academy of Sciences, Shanghai 201210, China
5. Center for Solid State Physics and New Materials, Institute of Physics, University of Belgrade, 11080 Belgrade, Serbia
6. Department of Physics, University of Colorado at Boulder, Boulder, CO 80309, USA
7. European X-ray Free-Electron Laser Facility, 22869 Schenefeld, Germany
8. Department Physik, Universität Siegen, 57072 Siegen, Germany

ABSTRACT

*By laser pump-probe time-resolved coherent magnetic X-ray diffraction imaging, we have measured the migration velocity of antiferromagnetic domain walls in the Mott insulator* $Sr_2IrO_4$ *at 100 K. Coherent diffraction patterns, recorded at the 106 pure-magnetic Bragg peak, showed fringes, which responded distinctly to the excitation laser. These patterns could not be inverted with current phasing algorithms, but could be modelled as four domains with phase shifts consistent with the antiferromagnetic structure. During the laser-induced demagnetization, we observe these domains and their domain walls moving at* $3\times10^6$ *m/s, significantly faster than acoustic velocities. This is understood to arise from a purely electronic spin contribution to the magnetic structure without any role for coupling to the crystal lattice.*

Magnetism is an important form of long-range order in condensed matter materials, which has a profound influence on their transport properties and a host of relevant applications. In the standard picture [1], magnetism arises from ordering of the electron spins in unpaired orbitals of the magnetic ions making up a material. An ordered antiferromagnetic (AFM) state forms when the spins are locally aligned in opposite directions with zero net magnetic moment [2]. In crystals, this results in a lower symmetry described by a supercell structure, leading to additional unique magnetic Bragg peaks in diffraction. The finite width of the magnetic reflections implies correlation over a finite distance in real space called the correlation length. This correlated region can be considered as a domain, separated from its neighbors by antiphase domain walls, which would be expected to thermally fluctuate on some time scale. Through the use of resonance with these electronic states, magnetic moments contribute to X-ray scattering, which can be used to study AFM structures [3,4].

The time scale for formation and destruction of magnetic domain structures is interesting and important for applications, such as magnetic memory devices. Because the spins are carried by the electrons, the domain dynamics might be expected to follow the time scale of electron diffusion inside a crystal, but can also be influenced by spin-lattice coupling which results in ionic displacements in a magnetic structure [5]. While electron-phonon coupling is widely measured to take place on a 1-5 ps time scale [6], its propagation into the crystal lattice travels at the speed of sound which can be much slower. Magnetic domains are subject to pinning at lattice defect sites, which can also slow their response time. Magnetic domain walls were reported to migrate at 172 m/s when measured by Ultrafast Lorenz Microscopy (UFM) in Ni thin films [7]. More recent UFM experiments on permalloy reported a spin-wave phase velocity of 1077 ± 34 m/s [8], suggesting it is limited by coupling to the lattice. Heat-assisted magnetic recording technology requires rapid magnetic domain switching [9]. Modelling of 5 nm ferromagnetic grains of FePt using the Landau-Lifshitz-Gilbert equations showed switching as fast as 10 ps [10]. In Gd–FeCo magnetic wires, domain-wall velocities of 1200 m/s have been reported [11], while laser driven AFM domain walls in $Sr_2Cu_3O_4Cl_2$ have recently been found to reach $5\times10^4$ m/s [12] and labyrinth domains in CoFe/Ni multilayers were seen to move at $6.6\times10^4$ m/s [13].

In this paper we interpret pump-probe coherent magnetic diffraction measurements made at an X-ray Free Electron Laser (XFEL) facility to address the question whether AFM domains and the domain walls separating them in $Sr_2IrO_4$ move at electronic, magnon or sound velocities, or otherwise. We report X-ray images of micron-sized clusters of magnetic phase domains which reconfigure within 100 fs following excitation by an optical laser, with their boundaries moving at $3\times10^6$ m/s. This is closer to a typical Fermi velocity, than the speed of sound, implying an electronic mechanism. The return relaxation is slower, taking longer than 10 ps, suggesting that lattice coupling may be involved. We also show that domains regrow in the same location every time.

A previous XFEL study showed sub-picosecond quenching of the antiferromagnetic (AFM) state in $Sr_2IrO_4$ followed by a rapid (< 2 ps) recovery of the local magnetic configuration seen in the Resonant Inelastic X-ray Scattering (RIXS) signal [14]. The long-range magnetic order was seen to be much slower, on the 100 ps time scale, independent of the local ionic configuration. Previous all-optical Kerr rotation studies similarly showed rapid demagnetization followed by a slower recovery [15]. None of the previously reported work was able to simultaneously record the positions of the domain walls, so could not report a spin velocity. Here, by use of ultra-fast coherent X-ray imaging, we are able to record the rate at which the domain-wall locations change and hence obtain their velocity in response to a pumping laser excitation.

We performed ultrafast magnetic Bragg Coherent Diffraction Imaging (BCDI) experiments on $Sr_2IrO_4$, carried out on the Materials Imaging and Dynamics (MID) instrument of the European XFEL facility [16]. The structure of $Sr_2IrO_4$, shown in Fig. 1(a), is tetragonal with a=b=5.499Å and c=25.784 Å [17]. The space group, $I4_1/acd$, has $4_1$ screw axes along the c-direction leading to the staggered pattern of 11° rotations of the $IrO_6$ octahedra shown. Below its Neel temperature, $T_N$=240K, $Sr_2IrO_4$ is antiferromagnetic (AFM), reported to have the AFM structure shown in Fig. 1(b), which breaks the 4-fold symmetry and leads to 90° magnetic twin domains with a slight orthorhombic distortion [19]. $Sr_2IrO_4$ shows exceptionally strong Spin-Orbit coupling [19], which causes the magnetic moments to be well-aligned with the octahedral rotations [20], which, in turn, leads to incomplete AFM cancellation within each $IrO_2$ layer of the unit cell. The 11° canting of the moments leads to a net moment along the y-axis shown as red arrows in Fig. 1(b). The net moments of the four $IrO_2$ layers in the unit cell alternate with the AFM ++-- pattern shown.  Important to this letter is that domains of the canted moments can also form, shifted by c/4. Fig. 1(c) shows a -++- domain, separated from the ++-- domain by a putative domain wall (red dashed line). Unlike the 90° twin domains [21] which have altogether different magnetic Bragg peaks [22], these “phase domains” and their “zero-angle” domain walls are not normally detected in AFM materials because their Bragg peak is at the same location in reciprocal space. However, they can be visualized using the coherent magnetic experiments we report here because their diffraction signals are sensitive to the phase produced by the different domains and their interference can be observed on the detector.

We previously demonstrated magnetic BCDI on $Sr_2IrO_4$ samples prepared in block shapes by Focused Ion Beam (FIB) methods [23]. In that previous magnetic BCDI work, a single magnetic domain was found to fill the FIB block of $1.2\times1.2\times1.2\ \mu m^3$ [23]. Previous work by scanning X-ray nanoprobe diffraction on larger crystals found side-by-side domains of both twin orientations up to 100 microns in size [21]. Correlation lengths obtained from high-resolution diffraction studies, which can be attributed to domain formation, were similarly large, 0.36 $\mu$m [25] and 1.2 $\mu$m [23].

We studied a large high-quality single crystal sample of $Sr_2IrO_4$, which was the parent crystal of the FIB-sample studied earlier [23]. The (106) magnetic reflection was pre-aligned by tilting the sample to the parallel (2 0 12) crystal reflection and offsetting the Bragg angle. Upon cooling to 100K and setting the XFEL self-seeding energy to precisely 11.215 keV, wavelength $\lambda = 0.1105$ nm, just below the Ir $L_3$ absorption edge [22], the speckled magnetic diffraction pattern shown in Fig. 2 was recorded on a Jungfrau detector with p = 75 x 75 $\mu m^2$ pixels, located D = 6 m from the sample. Compound Refractive Lens focusing optics were positioned 180 mm before the sample to give a calibrated focal spot size of 12 $\mu$m. A beam attenuator (transmitting 5%) was required to avoid ablation of material from the sample surface, observed when too fine a focus or too little attenuation were used. The speckled diffraction arises from coherent interference between the naturally occurring AFM domains within the beam size. By searching the position of the sample in the beam, we were able to locate isolated magnetic domains, perhaps separated by twin domains with the other orthorhombic orientation [19], that would have Bragg peaks at (016). The diffraction patterns were found to be quasi two-dimensional (2D) in the sense that they did not evolve strongly with small changes of the sample rocking angle.

The XFEL beam is highly coherent and the diffraction patterns were slightly different for each {106} domain investigated. The example shown in Fig. 2(a), was chosen for further investigation under laser excitation. It is strongly elongated on the area detector close to a 45° angle, which is understood to come from the 7.35° grazing incidence in the tilted sample geometry, shown in Supplementary Figure S1. The observed pattern has two broad central maxima and regular side-fringes, which is loosely interpreted as due to a pair of side-by-side antiphase magnetic domains, causing a cancellation at the center. The fringes are 100 pixels long, with a fringe-to-fringe spacing of 6 pixels. In this work we consider the images in the coordinate system of the detector, in which the real-space pixel size is $\lambda$D/Np, where *N* is the array size of the discrete Fourier transform used. We can therefore estimate the domains to be about 100 nm wide and 1.5 $\mu$m long. Two side-by-side antiphase domains, one micron long, would explain the symmetric splitting of the peak center and the side fringes (see below).

The temporal evolution of the diffraction pattern following optical excitation was recorded in an optical-pump X-ray probe scheme, by means of pump-probe delay scans. The 50 fs optical laser pulses centered around 800 nm (1.55 eV) impinged on the sample nearly colinear with the X-ray beam. This energy is above the charge gap of $Sr_2IrO_4$, so It drives excitation between the $J_{eff}$=1/2 and $J_{eff}$=3/2 bands, creating holon-doublons propagating through the lattice and destroying magnetic order within a few 100 fs [14, 15, 26]. The incident fluence was 1.4 mJ/cm$^2$, sufficient to cause a 23% drop in the overall (106) peak diffraction intensity. The diffraction pattern shown in Fig. 2(a) was averaged over all negative time delays, $\Delta t$, while the one in Fig. 2(b) is averaged over $0 < \Delta t < 1$ ps. The two-time correlation function [27] of the diffraction, as a function of time delay is shown in Fig. 2(c). Despite some fluctuation of the correlations both at negative and positive time delays believed to come from imperfect normalization of beam monitoring, clear changes in the diffraction pattern are seen, with a sudden jump in correlation at $\Delta t \approx 0$. The correlation drops to 0.93 at positive $\Delta t$ and relaxes afterwards. Careful inspection of Figs. 2(a) and (b) show the fringe spacing increases slightly for positive $\Delta t$.

We attempted to obtain real-space images of the domains by inversion of the diffraction patterns using 500 cycles of the Hybrid Input-Output (HIO) method alternating with error reduction and a fixed elliptical “support” estimated from the speckle shape in the diffraction pattern [28,29] and also using the guided HIO approach [30]. However, it was not possible to obtain a unique solution from the 2D diffraction pattern, perhaps due to the low signal-to-noise ratio. Instead, we obtained a number of similar solutions agreeing with the data; this diversity of solutions is considered to represent the propagation of noise in the raw data, arising from limited counting statistics, giving agreement $0.048 < \chi^2 < 0.063$ [31]. Typical examples of the reconstructed phase images, shown in Fig. 3(a) for HIO and Fig. 3(b) for guided HIO, have phase ramps and offsets removed. They typically show 4 to 6 regions of flat phase with steps between resembling domains and discrete phases distributed over $-\pi < \phi < \pi$. This block structure suggests the presence of AFM domains shifted by fractions of a unit cell along the c-axis separated by domain walls as illustrated in Fig.1. The micron-sized images seen in the reconstructions in Fig. 3, consistent with the speckle size, are smaller than the nominal X-ray beam size, so might indicate the magnetic domains are surrounded by invisible regions with the other orthorhombic distortion [21].

The poor reproducibility of the reconstructions made it impossible to track the small changes due to the laser excitation of the AFM structure. So, an explicit, simplified 4-domain model, shown in Fig. 3(c), was introduced and manually adjusted to give good qualitative agreement with the diffraction data. This model has two axis dimensions of the ellipse and one rotation, three independent phase values (since the fourth can arbitrarily set to zero) and six parameters to position the domain boundaries. The elliptical shape was chosen for simplicity. We used a Voronoi construction [32] where the phase is assigned as belonging to the domains with the closest seed position, expressed as a fraction of the ellipse axis lengths. This model, with 12 adjustable parameters, was then optimized for best agreement with the diffraction pattern using the Powell scipy.optimize.minimize algorithm [33], shown in Table 1 and Fig. 3(c,d). The agreement with the diffraction data is shown in Supplementary Figure S2.

The diffraction patterns measured at different pump-probe delay times were then fitted to the 12-parameter model, resulting in a statistically good fit with a correlation coefficient of 0.92.

The fit was stable even for single frames of 100 fs, but is plotted in Fig. 4 for steps of 200 fs after binning frames. The binning reduced the error bar, estimated from the variations of the negative time-delay fit parameters. As seen in Table 1, only one of the 12 model parameters was found to vary significantly with the laser time delay, $\Delta t$, which is the length of the elliptical envelope containing the domains, $h(\Delta t)$. Supplementary Figure S3 shows the variations of the other 11 parameter of the model with $\Delta t$. Fitting the trend to a double exponential decay in Fig. 4,

$$h(\Delta t) = h_0 \times e^{-\frac{\Delta t - t_0}{\tau_1}} \times \left(1 - e^{-\frac{\Delta t - t_0}{\tau_2}}\right) + const \quad (1)$$

yielded a 15% drop of $h(\Delta t)$ in $\tau_1 = 40 \pm 30$ fs and a recovery time of $\tau_2 = 20$ ps, roughly consistent with the trends in the diffraction signal seen in the correlation coefficient of the raw data, as discussed above.

| | -5 ps < $\Delta t$ < 0 | 0 < $\Delta t$ < +5 ps |
|---|---|---|
| Major half-axis ($\mu$m) | 1.017 | 0.890 |
| Minor half-axis ($\mu$m) | 0.281 | 0.307 |
| Rotation (radians) | 0.673 | 0.689 |
| Voronoi 1 (fractions) | (0.106, -0.910) | (0.132, -0.642) |
| Voronoi 2 (fractions) | (0.034, 0.671) | (0.107, 0.714) |
| Voronoi 3 (fractions) | (0.266, -0.026) | (0.25, 0.053) |
| Voronoi 4 (fractions) | (0,0) | (0,0) |
| Phase 1 (radians) | -1.874 | -1.90 |
| Phase 2 (radians) | 1.412 | 1.327 |
| Phase 3 (radians) | 2.819 | 2.858 |
| Phase 4 (radians) | 0.0 | 0.0 |

Table 1. Fit parameters of the 4-domain model of magnetic domains in the detector coordinate system, shown in Fig. 3. The pump-probe diffraction data were averaged into the two time ranges indicated before fitting the model. The first three parameters define the elliptical shape of the area filled with domains. The Voronoi parameters are the fitted domain centers, as fractions of the (minor, major) axis lengths. The phases are the fitted values relative to the fourth domain, fixed to be zero at position (0,0).

The double exponential response of the diffraction intensity to the laser excitation has been reported for $Sr_2IrO_4$ before. Dean et. al. [14] reported $\tau_1 = 268 \pm 16$ fs and $\tau_2 = 197$ ps at a fluence of 2.67 mJ/cm$^2$, similar to that used in our experiment, in pump-probe time-resolved magnetic scattering at the (3 2 28) peak. Laser reflectivity measurements [15] reported $\tau_1 = 262$ fs and $\tau_2 = 0.72$ ps and demagnetization by Kerr rotation $\tau_1 = 345$ fs and $\tau_2 = 1.52$ ps, both at a sample temperature of 80 K and a fluence of 0.1 mJ/cm$^2$. Since both studies reported strong increases of $\tau_2$ with fluence, we consider them both to agree roughly with our results for the demagnetization and recovery times, given the large uncertainties in the trend fitting in Fig. 4 and the different pump wavelength used in Ref. [14].

To support the validity of the real-space image modelling of the changes in the observed diffraction pattern, we also performed a direct fit to the data. We integrated the diffraction pattern in Fig 2 along the diagonal direction to extract its intensity profile, shown as amplitude (square-root) in Fig 4(b). This profile was least-squares fitted to the Fourier transform of a 1D block of scattering material with an antiphase domain boundary at the center,

$$
\begin{aligned}
A(q) &= \int_{-a}^{a} e^{i\phi(x)} e^{iqx}\, dx, \quad \phi(x) = \begin{cases} \pi, & x < 0 \\ 0, & x > 0 \end{cases} \\
&= \int_{0}^{a} e^{iqx} dx - \int_{-a}^{0} e^{iqx} dx = \frac{2}{iq}\big(\cos(qa) - 1\big),
\end{aligned}
\tag{2}
$$

where *2a* is the length of the block representing the magnetic domain in 1D. This *A(q)* function resembles a *sinc(qa)* function, except for having a node at the center coming from the phase step. The fit curves in Fig 4(b) show a good agreement of this simpler 1D model both for the $\Delta t < 0$ and $\Delta t > 0$ pump-probe delay ranges, where the frequency of the side fringes can clearly be seen to change. This 1D description is simple enough to allow reliable fitting of single 100 fs-spaced frames of the $\Delta t$ time series, for which the fitted block half-length, *a* in Eq(2), is plotted in Fig 4(c). The trend is very similar to that of the 4-domain model in Fig. 4(a) and the time constants were $\tau_1$ = 87 fs and $\tau_2$ = 19.6 ps showing good agreement between the two descriptions and supporting the validity of the 4-domain picture used above.

While the sub-picosecond excitation of the magnetic structure has been seen before, we have now visualized how the associated spatial magnetic arrangement changes too. Within the $\tau_1 = 40\pm30$ fs excitation, the outer boundary of the domains is seen to move by 0.127 μm and the internal domain walls by a fraction of that distance in Fig. 3(e, f). The block half-length of the second model in Fig. 4(c) contracts by $\Delta a$ = 0.18 μm in 87 fs. This represents a domain velocity of $3\times10^6$ and $2.1\times10^6$ m/s for the two models, typical of an electronic Fermi velocity and far faster than the speed of sound. We conclude there is no acoustic component associated with the excitation, for example a change in rotation angle of the octahedra, which would be limited to sound velocities. The magnetic structure in Fig. 1 is purely composed of spins, so the velocity we measure is a spin velocity. Since the spins are electronic, it is reasonable to expect their migration to be at a similar velocity.

In conclusion, we have imaged the dynamics of a cluster of antiferromagnetic domains in $Sr_2IrO_4$ excited by a femtosecond optical laser by magnetic Bragg coherent diffractive imaging. We observe a distinct contraction of the domain structure, including its internal domain walls, with a fitted time constant of $40\pm30$ fs. Such "breathing" behavior has been seen before in thermally-driven magnetic domain fluctuations [34]. The corresponding domain wall migration velocity is $3\times10^6$ m/s, which is typical of an electronic Fermi velocity. This informs us there is no role of electron-lattice coupling in the domain wall motion, instead involving a pure flipping of spins.

Acknowledgements: We acknowledge European XFEL in Schenefeld, Germany, for the provision of X-ray free-electron laser beamtime at the MID (Materials Imaging and Dynamics) instrument under proposal numbers p3331 and p6156 [35]. Work at Brookhaven National Laboratory was supported by the U.S. Department of Energy, Office of Science, Office of Basic Energy Sciences, under Contract No. DE-SC0012704. Work at Argonne National Laboratory was supported by the U. S. Department of Energy, Office of Science, Office of Basic Energy

Sciences, under Contract No. DE-AC02-06CH11357. Work performed at UCL was supported by EPSRC. Work at Shanghai Advanced Research Institute was funded by the '100 Talents Project' of the Chinese Academy of Sciences. Work at U. Colorado acknowledges NSF support via Grant No. DMR 2204811. All relevant data and processing scripts are published online [35, 36].

Justification Paragraph (75 words): While laser-driven demagnetization and magnetic X-ray scattering in $Sr_2IrO_4$ are both well-established, the combination in a pump-probe experiment at an X-ray free electron laser is new. Micron-sized antiferromagnetic domains have been seen by other methods, but not the more subtle phase domains to which coherent scattering is sensitive. By measuring the temporal and spatial changes in the domain walls in a single experiment, we are able to establish a wall-migration velocity for the first time.

1. L. D. Landau, A possible explanation of the field dependence of the susceptibility at low temperatures, Phys. Z. Sowjet, 4, 675 (1933)
2. M. L. Néel, Propriétées magnétiques des ferrites; Férrimagnétisme et antiferromagnétisme, Annales de Physique 12 137 (1948)
3. M. Blume and D. Gibbs, Polarization dependence of magnetic x-ray scattering, Phys. Rev. B 37 1779 (1988)
4. U. Staub, G. I. Meijer, F. Fauth, R. Allenspach, J. G. Bednorz, J. Karpinski, S. M. Kazakov, L. Paolasini and F. d'Acapito, Direct Observation of Charge Order in an Epitaxial $NdNiO_3$ Film, Phys. Rev. Lett. 88 126402 (2002)
5. P. B. Allen, Theory of thermal relaxation of electrons in metals, Phys. Rev. Lett. 59 1460 (1987)
6. J. K. Chen, D. Y. Tzou, J. E. Beraun, A semiclassical two-temperature model for ultrafast laser heating, Int. J. Heat Transfer 49 307 (2006)
7. H. S. Park, J. S. Baskin, and A. H. Zewail, 4D Lorentz Electron Microscopy Imaging: Magnetic Domain Wall Nucleation, Reversal, and Wave Velocity, Nano Lett. 10 3796 (2010)
8. C. Liu, F. Ai, S. Reisbick, A. Zong, A. Pofelski, M-G. Han, F. Camino, C. Jing, V. Lomakin and Y. Zhu, Correlated spin-wave generation and domain-wall oscillation in a topologically textured magnetic film, Nature Materials 24 406 (2025)
9. D. Weller, G. Parker, O. Mosendz, A. Lyberatos, D. Mitin, N. Y. Safonova and M. Albrecht, FePt heat assisted magnetic recording media, J. Vac. Sci. Technol. B 34 060801 (2016)
10. M. O. A. Ellis and R. W. Chantrell, Switching times of nanoscale FePt: Finite size effects on the linear reversal mechanism, Appl. Phys. Lett. 106 162407 (2015)
11. S. Ranjbar, S. Kambe, S. Sumi, P. V. Thach, Y. Nakatani, K. Tanabea and H. Awano, Elucidation of the mechanism for maintaining ultrafast domain wall mobility over a wide temperature range, Mater. Adv. 3 7028 (2022)
12. K. L. Seyler, H. Zhang, D. Van Beveren, C. R. Rotundu, Y. S. Lee, R. Cheng and D. Hsieh, High-speed antiferromagnetic domain walls driven by coherent spin waves, Nat. Commun. 16 9836 (2025)
13. R. Jangid, N. Z. Hagström, M. Madhavi, K. Rockwell, J. M. Shaw, J. A. Brock, M. Pancaldi, D. De Angelis, F. Capotondi, E. Pedersoli, H. T. Nembach, M. W. Keller, S. Bonetti, E. E. Fullerton, E. Iacocca, R. Kukreja and T. J. Silva, Extreme Domain Wall Speeds under Ultrafast Optical Excitation, Phys. Rev. Lett. 131 256702 (2023)
14. M. P. M. Dean, Y. Cao, X. Liu, S. Wall, D. Zhu, R. Mankowsky, V. Thampy, X. M. Chen, J. G. Vale, D. Casa, J. Kim, A. H. Said, P. Juhas, R. Alonso-Mori, J. M. Glownia, A. Robert, J.

Robinson, M. Sikorski, S. Song, M. Kozina, H. Lemke, L. Patthey, S. Owada, T. Katayama, M. Yabashi, Yoshikazu Tanaka, T. Togashi, J. Liu, C. Rayan Serrao, B. J. Kim, L. Huber, C.-L. Chang, D. F. McMorrow, M. Först and J. P. Hill, Ultrafast energy- and momentum-resolved dynamics of magnetic correlations in the photo-doped Mott insulator $Sr_2IrO_4$, Nat. Mater. 15 601 (2016)

15. D. Afanasiev, A. Gatilova, D. J. Groenendijk, B. A. Ivanov, M. Gibert, S. Gariglio, J. Mentink, J. Li, N. Dasari, M. Eckstein, T. Rasing, A. D. Caviglia and A. V. Kimel, Ultrafast Spin Dynamics in Photodoped Spin-Orbit Mott Insulator $Sr_2IrO_4$, Phys Rev X 9 021020 (2019)
16. A. Madsen, J. Hallmann, G. Ansaldi, T. Roth, W. Lu, C. Kim, U. Boesenberg, A. Zozulya, J. Moeller, R. Shayduk, M. Scholz, A. Bartmann, A. Schmidt, I. Lobato, K. Sukharnikov, M. Reiser, K. Kazarian and I. Petrov, Materials Imaging and Dynamics (MID) instrument at the European X-ray Free-Electron Laser Facility, J. Synchrotron Rad. 28 637 (2021)
17. F. Ye, S. Chi, B. C. Chakoumakos, J. A. Fernandez-Baca, T. Qi and G. Cao, Magnetic and crystal structure of $Sr_2IrO_4$: a neutron difffraction study, Phys. Rev. B 87, 140406 (2013)
18. K. Momma and F. Izumi, VESTA 3 for three-dimensional visualization of crystal, volumetric and morphology data, J. Appl. Cryst. 44 1272 (2011)
19. B. J. Kim, H. Jin, S. J. Moon, J-Y. Kim, B-G. Park, C. S. Leem, J. Yu, T. W. Noh, C. Kim, S-J. Oh, J-H. Park, V. Durairaj, G. Cao and E. Rotenberg, Novel $J_{eff}$=1/2 Mott State Induced by Relativistic Spin-Orbit Coupling in $Sr_2IrO_4$, Phys. Rev. Lett. 101 076402 (2008)
20. S. Boseggia, H. C. Walker, J. Vale, R. Springell, Z. Feng, R. S. Perry, M. Moretti Sala, H. M. Rønnow, S. P. Collins and D. F. McMorrow, Locking of iridium magnetic moments to the correlated rotation of oxygen octahedra in $Sr_2IrO_4$ revealed by x-ray resonant scattering, J. Phys. Condens. Matter 25 422202 (2013)
21. T. Choi, Z. Zhang, H. Kim, S. Park, J-W. Kim, K. J. Lee, Z. Islam, U. Welp, S. H. Chang and B. J. Kim, Nanoscale Antiferromagnetic Domain Imaging using Full-Field Resonant X-ray Magnetic Diffraction Microscopy, Adv. Mater. 34 2200639 (2022)
22. B. J. Kim, H. Ohsumi, T. Komesu, S. Sakai, T. Morita, H. Takagi and T. Arima, Phase-Sensitive Observation of a Spin-Orbital Mott State in $Sr_2IrO_4$, Science 323 1329 (2009)
23. L. Wu, W. Wang, T. A. Assefa, A. F. Suzana, J. Diao, G. Cao, R. J. Harder, W. Cha, K. Kisslinger, M. P. M. Dean and I. K. Robinson, Anisotropy of Antiferromagnetic Domains in a Spin-orbit Mott Insulator, Physical Review B 108 L020403 (2023)
24. T. Choi, Z. Zhang, H. Kim, S. Park, J-W. Kim, K. J. Lee, Z. Islam, U. Welp, S. H. Chang and B. J. Kim, Nanoscale antiferromagnetic domain imaging using full-field resonant x-ray magnetic diffraction microscopy, Adv. Mater., 34, 2200639 (2022)
25. J-W. Kim, S. H. Chun, Y. Choi , B. J. Kim, M. H. Upton and P. J. Ryan, Controlling symmetry of spin-orbit entangled pseudospin state through uniaxial strain, Phys. Rev. B 102 054420 (2020)
26. D. G. Mazzone, D. Meyers, Y. Cao, J. Vale, C. Dashwood, A. J. A. James, N. J. Robinson, J. Q. Lin, V. Thampy, Y. Tanaka, A. Johnson, H. Miao, R. Wang, T. A. Assefa, J. Kim, D. Casa, R. Mankowsky, D. Zhu, R. Alonso-Mori, S. Song, H. Yavas, T. Katayama, M. Yabashi, Y. Kubota, S. Owada, J. Liu, J. Yang, Y. Shi, R. M. Konik, I. K. Robinson, J. P. Hill, D. F. McMorrow, M. Foerst, S. Wall, X. Liu and M. P. M. Dean, Laser-Induced Transient Magnons in $Sr_3Ir_2O_7$ Throughout the Brillouin Zone, Proceedings of the National Academy of Sciences USA, 118 e2103696118 (2021)
27. O. G. Shpyrko, X-ray photon correlation spectroscopy, J. Synchrotron Rad. 21, 1057–1064 (2014)

28. J. R. Fienup, Reconstruction of an object from the modulus of its Fourier transform, Opt. Lett. 3 27 (1978)
29. I. Robinson and R. Harder, Coherent X-ray diffraction imaging of strain at the nanoscale, Nat. Mater. 8, 291 (2009)
30. C. C. Chen, J. Miao, C. W. Wang, and T. K. Lee, Application of optimization technique to noncrystalline x-ray diffraction microscopy: Guided hybrid input-output method, Phys. Rev. B 76, 064113 (2007)
31. R. H. T. Bates, Global solution to the scalar inverse scattering problem, J. Physics A: Mathematics & General, L80-L82 (1975)
32. G. Voronoi, Nouvelles applications des paramètres continus à la théorie des formes quadratiques. Journal für die Reine und Angewandte Mathematik 133 97 (1908)
33. M. J. D. Powell, An efficient method for finding the minimum of a function of several variables without calculating derivatives, The Computer Journal 7 155 (1964)
34. L. Wu, Y. Shen, A. M. Barbour, W. Wang, D. Prabhakaran, A. T. Boothroyd, C. Mazzoli, J. M. Tranquada, M. P. M. Dean and I. K. Robinson, Real Space imaging of Spin Stripe Domain Fluctuations in a Complex Oxide, Phys. Rev. Lett. 127 275301 (2021)
35. Data recorded at the experiment at the European XFEL are available at doi: 10.22003/XFEL.EU-DATA-006156-00
36. Doi: 10.5281/zenodo.17575334

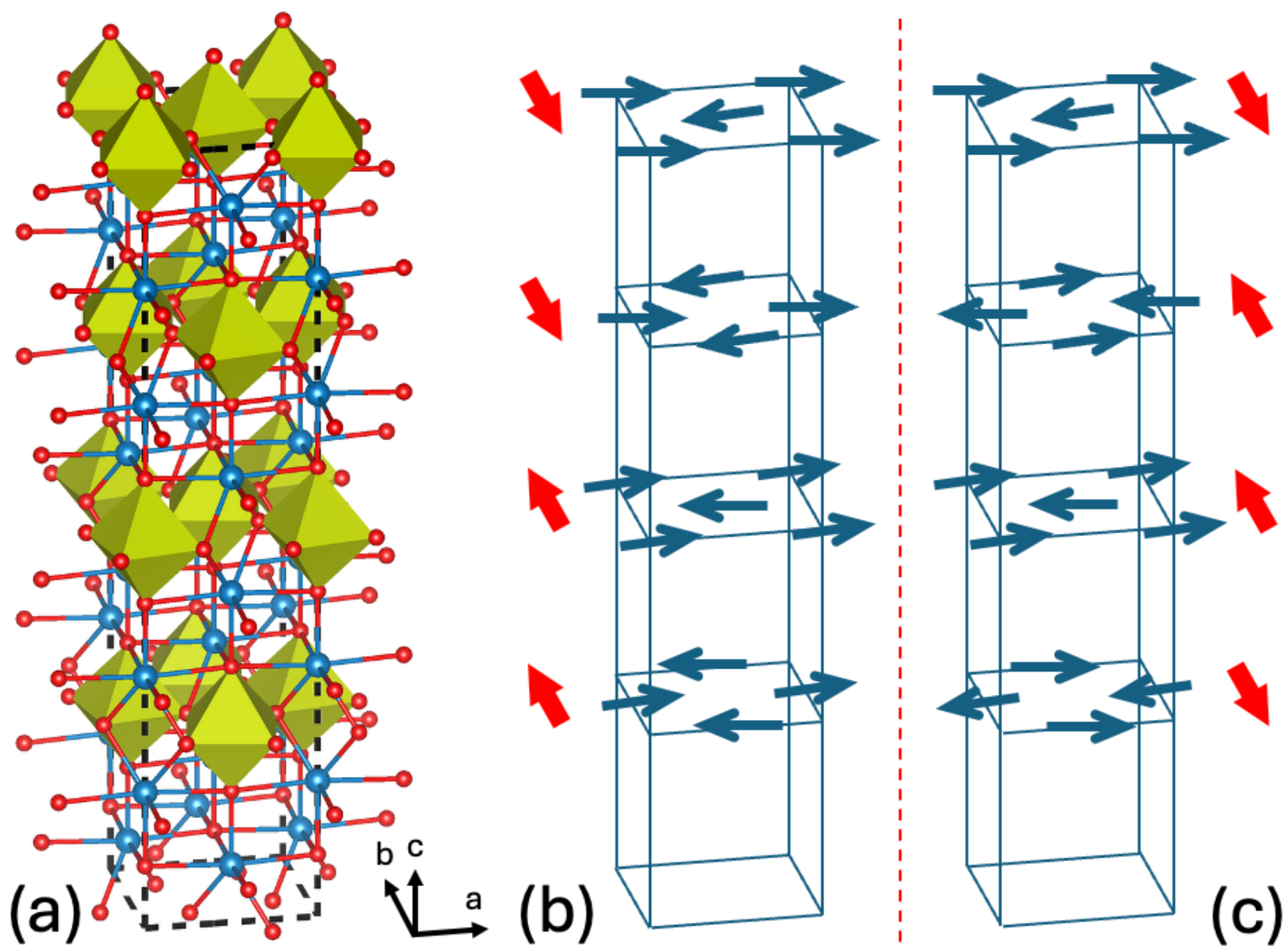

Figure 1: (a) Atomic structure of strontium iridate, $Sr_2IrO_4$. $Sr^{2+}$ ions are blue, $O^{2-}$ ions are red and the $Ir^{4+}$ ions are at centers of the canted octahedra. (b,c) Magnetic structure showing two of the four possible phase domains (++-- and -++- respectively) with an indicative domain wall (dashed line) between them. The red arrows indicate the net in-plane $Ir^{4+}$ magnetic moments due to canting of the octahedra in the structure. The figure was prepared with the use of the VESTA imaging software [18].

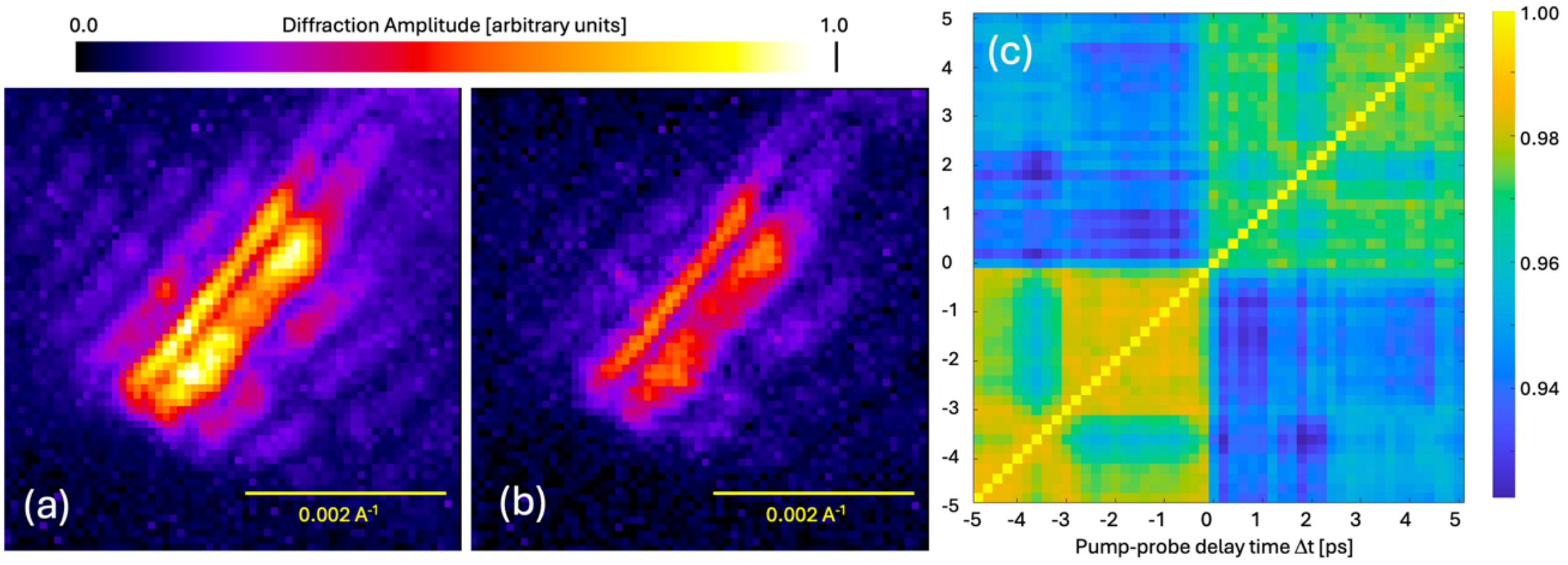

Figure 2. Amplitude (square root of the measured intensity) of the coherent diffraction pattern around the 106 magnetic Bragg peak of $Sr_2IrO_4$ recorded on the Jungfrau detector (a) before and (b) over the first picosecond after the laser excitation. (c) Two-time correlation function between all measured diffraction patterns, binned into 200 fs long groups.

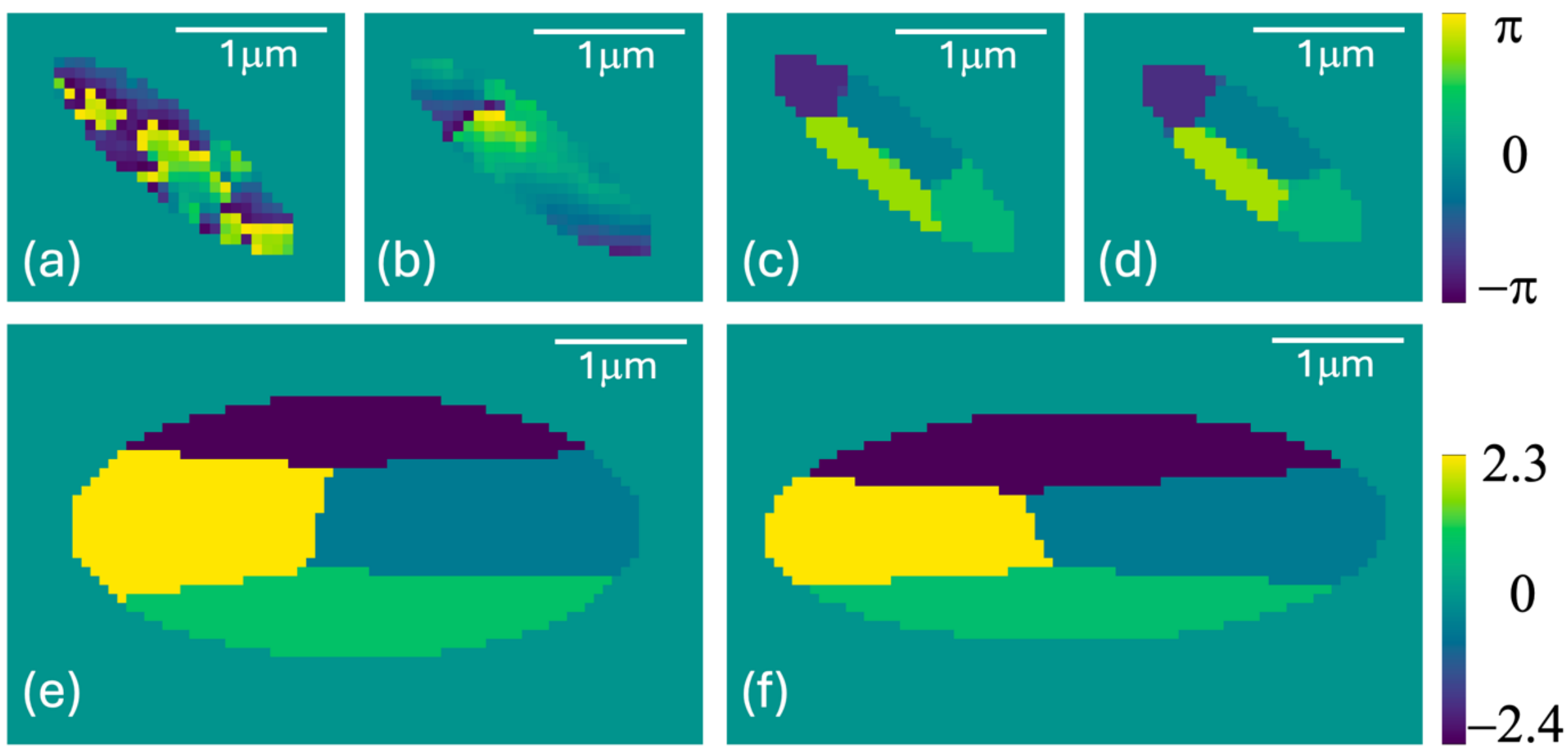

Figure 3: (a) Phase reconstruction of the observed diffraction pattern at negative time delay, -5<Δt<0 ps, in a fixed support using (a) 500 cycles of alternating HIO_ER (b) guided HIO [30] with 5 generations of 15 populations (c) the 12-parameter 4-domain model described in the text. (a) and (b) are typical examples of a wide range of non-reproducible structures, shown in the "detector" coordinate frame, the Fourier transform of the data shown in Fig.2. (d) the same model fitted to positive time-delay data, 0<Δt<5 ps. (e) and (f) are the same as (c) and (d), but transformed to the view on the 001 face of the sample.

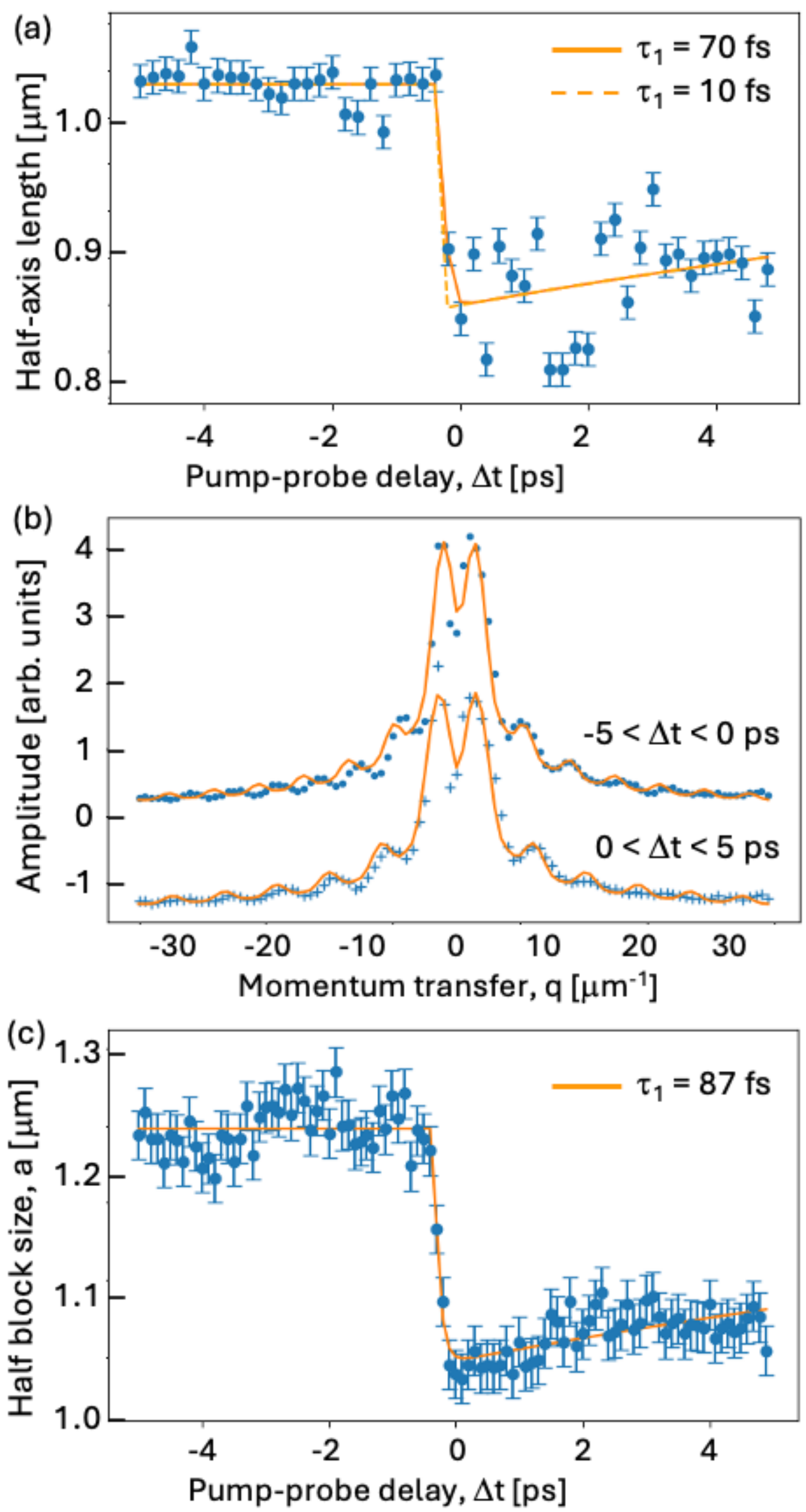

Figure 4(a) Time dependence of the domain envelope length parameter of the 4-domain model used to fit the diffraction pattern. The trend has been fit to a double exponential function with time constant $\tau_1$ for the excitation and $\tau_2$ for its relaxation. Two fit curves with $\tau_1$ = 10 fs and 70 fs indicate the range of possible fits. (b) Direct fit to the raw magnetic diffraction amplitude data integrated along a 38-degree diagonal direction over the time ranges -5 ps < $\Delta t$ < 0 (top) and 0 < $\Delta t$ < +5 ps (bottom, offset for clarity). The fitting function was Eq(2) quadratically added to a Lorentzian background function. (c) Values of the block half-length parameter *a* in Eq(2) extracted for single XFEL shots and fit to the same double exponential.

Supplementary Figures:

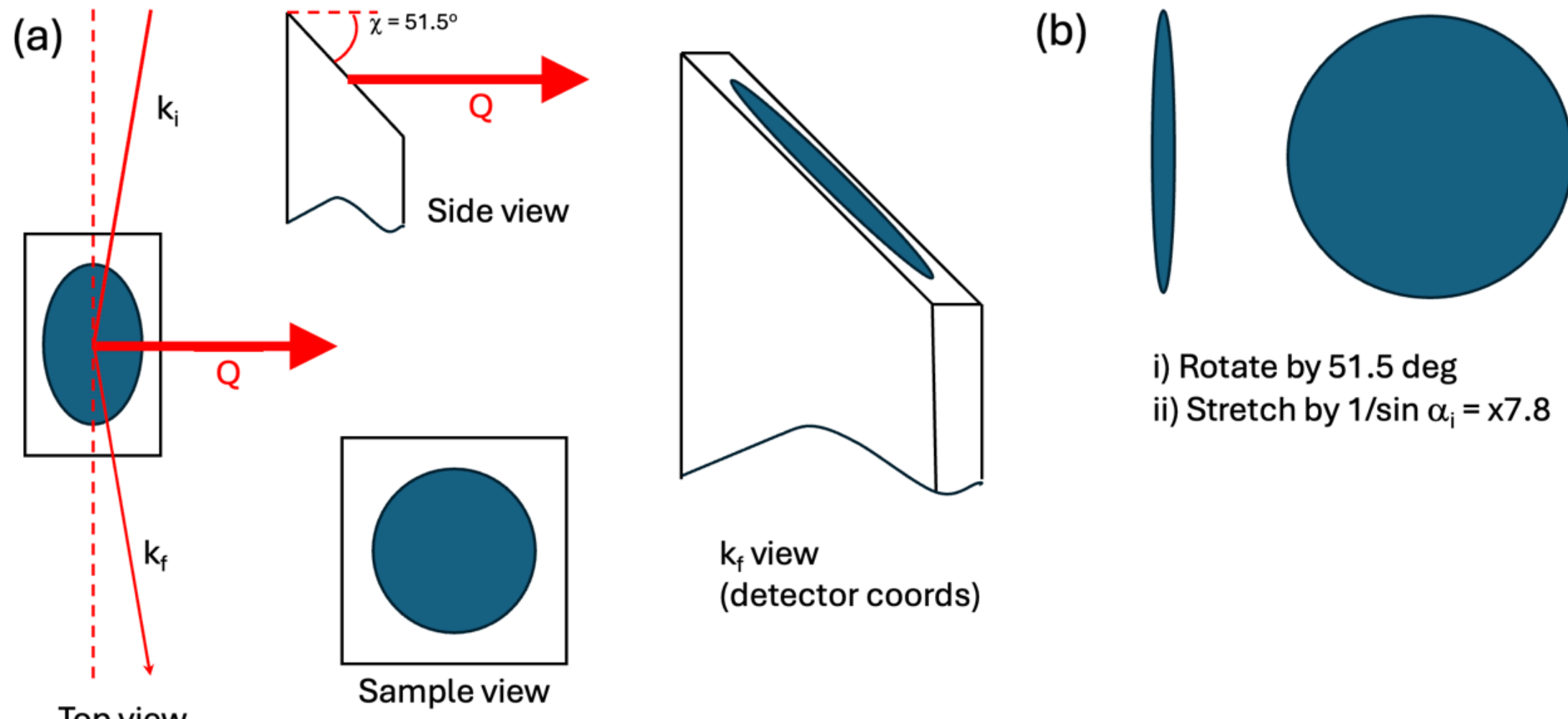

Figure S1: (a) Diffraction geometry of the X-ray setup at the MID instrument of XFEL. The Q-vector lies in the horizontal plane, and the sample face is tilted by 51.5° to reach the 106 reflection (b) coordinate transformation required to convert the reconstructed image from the detector view into the surface-normal view.

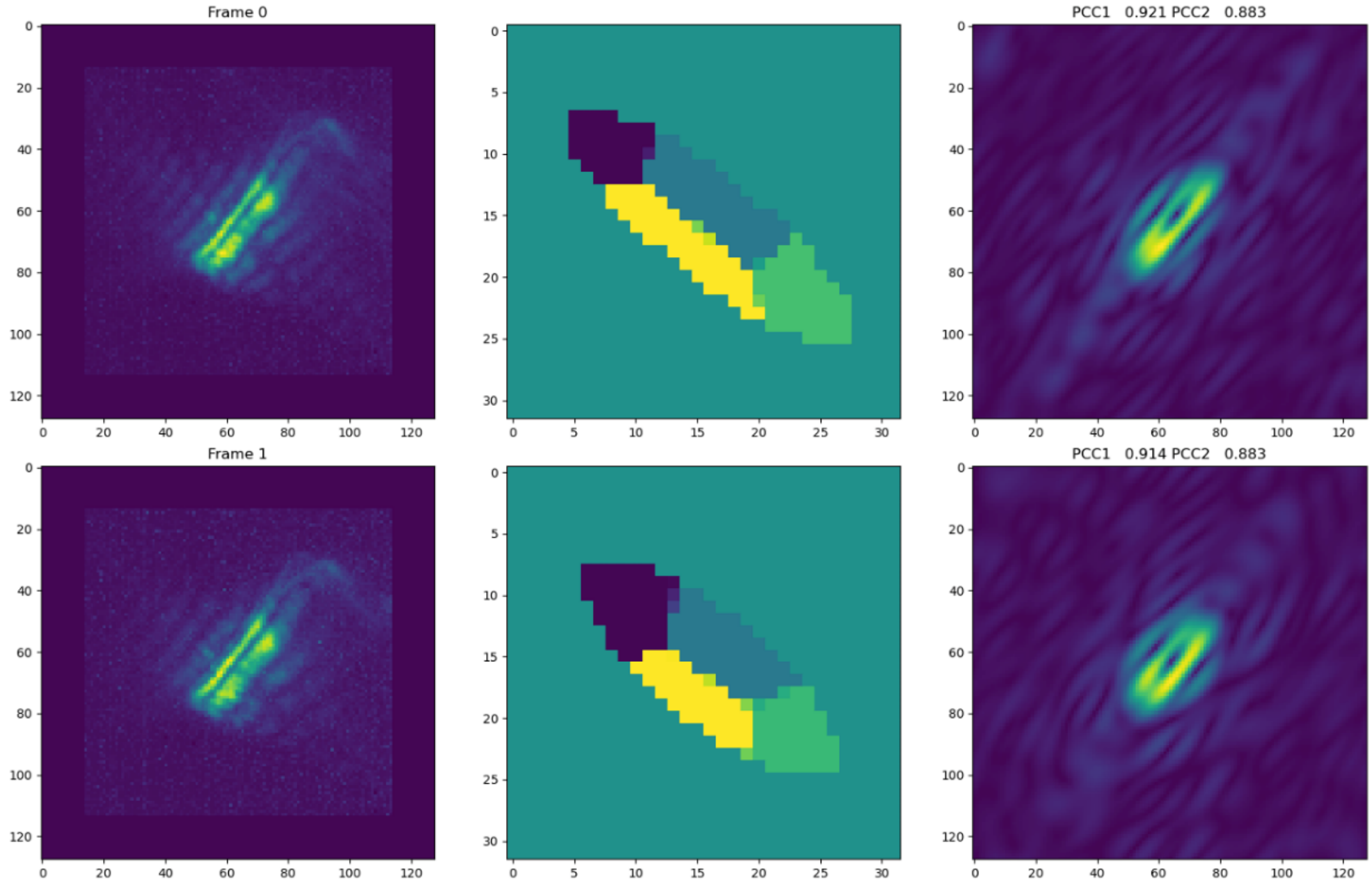

Figure S2: Fitting of the 4-domain model to the XFEL magnetic coherent diffraction data summed over ranges -5 ps < $\Delta t$ < 0 (frame 0, top) and 0 < $\Delta t$ < +5 ps (frame 1, bottom). The first column is the measured diffraction amplitude, the second column is the fit model colored according to the values of the phases and the third column is the best-fitting diffraction amplitude distribution.

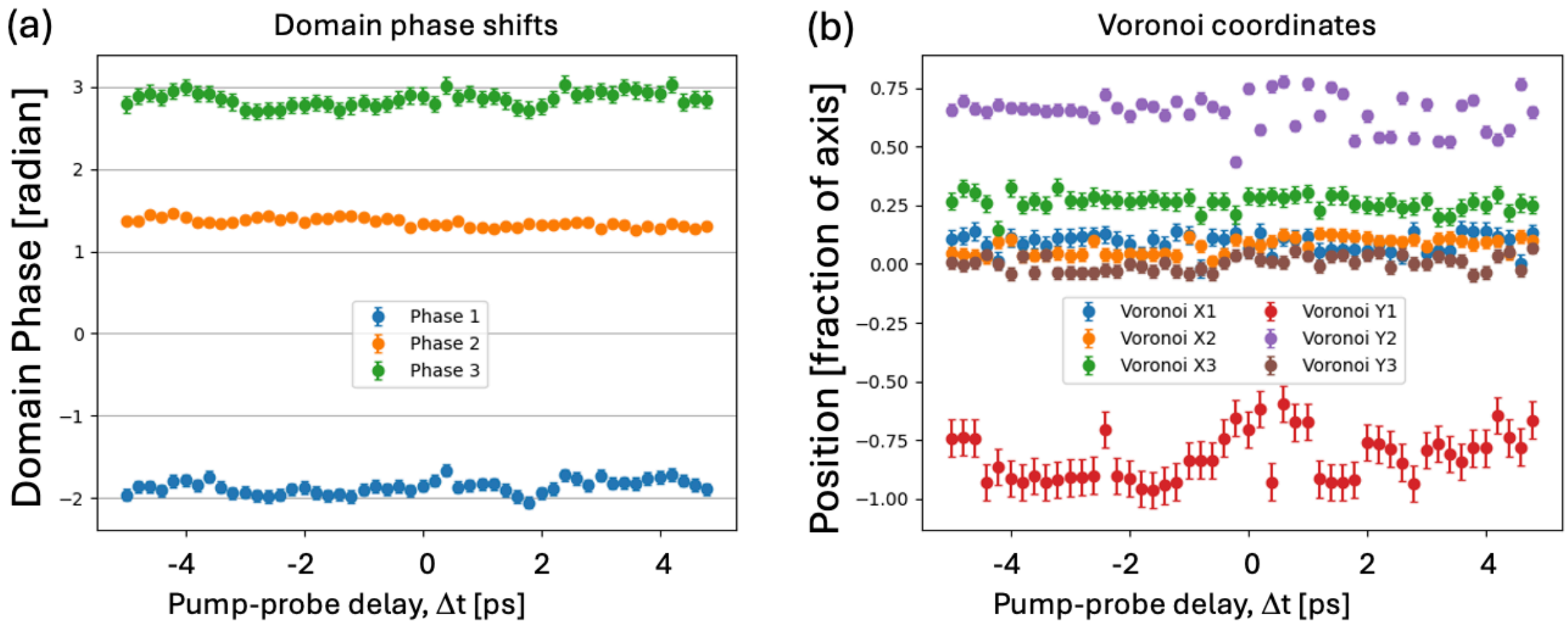

Figure S3: Time dependence of the other parameters of the model listed in Table 1, used to fit the diffraction pattern. (a) Domain phase shifts. (b) Voronoi (x,y) coordinates as a fraction of the axis length of the ellipse, as listed in Table 1. The fourth domain is fixed at (0,0) with a phase of 0.